\documentstyle[12pt,iopconf,graphicx,amsmath,amssymb]{article}

\begin{document}
\title{Linear and Nonlinear Dynamics of Relativistic Tori}

\author{Luciano Rezzolla}

\affil{SISSA, International School for Advanced Studies, Trieste, Italy}

\affil{INFN, Sezione di Trieste, Trieste, Italy}

\beginabstract 
	I present results of two-dimensional general relativistic
	hydrodynamical simulations of constant specific angular momentum
	tori orbiting around a Schwarzschild black hole.  After
	introducing axisymmetric perturbations, these objects either
	become unstable to the runaway instability or respond with
	regular oscillations. The latter, in particular, are responsible
	for quasi-periodic bursts of accretion onto the black hole as
	well as for the emission of intense gravitational radiation, with
	signal-to-noise ratios at the detector which are comparable or
	even larger than the typical ones expected in stellar-core
	collapse.  
\endabstract

%
%=======================================================
\section{Introduction}
\label{intro}
%=======================================================

	Stationary barotropic configurations around black holes in which
the matter moves in circular orbits within isobaric equipotential
surfaces have been studied for more than twenty years~\cite{abramowicz78,
kozlowski78}. An important feature of these non-Keplerian solutions is
that the equipotential surfaces possess, in general, a sharp cusp on the
equatorial plane, which behaves as an effective Lagrangian point, thus
providing accretion even in the absence of a shear viscosity in the
fluid.  The dynamical response of these relativistic tori to
perturbations has important astrophysical implications and will be
discussed here in connection with the following three issues.

	Firstly, the stability with respect to accretion of matter
through the cusp. Any fluid element orbiting at the cusp will, upon
experiencing the smallest inward radial perturbation, accrete onto the
black hole. This will change the black hole mass, affect the stationary
solution and, most importantly, the location of the cusp. The dynamics of
the cusp can then have two possible evolutions. If, as a result of the
newly accreted matter, the cusp moves to {\it smaller} radial positions,
all fluid elements will be on {\it stable} circular orbits and no further
accretion will be possible. If, on the other hand, the cusp moves to {\it
larger} radial positions faster than the inner edge of the torus (which
has shrunk as a result of the mass loss), the new configuration will {\it
not} be of stable equilibrium and additional material can therefore
accrete onto the black hole. The instability which is produced in this
way is then referred to as the ``runaway'' instability
\cite{abramowicz83} and could have important implications on current
models of $\gamma$-ray bursts~\cite{daigne97, meszaros02}.

	Secondly, the possibility of calculating numerically the linear
and nonlinear hydrodynamical evolution of the torus provides a useful
tool for relativistic disco-seismology in two dimensions.  Having an
internal structure governed by the balance of gravitational sources,
pressure gradients and centrifugal forces, these relativistic tori
possess non-trivial oscillation properties which are still essentially
unexplored. By introducing perturbations, it is possible to excite the
relevant modes of oscillations and calculate the characteristic
eigenfrequencies to high accuracy and even in the absence of a
perturbative analysis. Furthermore, during each of these oscillations a
significant amount of matter could fall onto the black hole and, in a
realistic scenario, it is reasonable to expect that before this matter
reaches the event horizon, it will loose part of its binding energy,
increase its temperature and emit electromagnetic radiation. For this
reason, the quasi-periodic accretion produced in this way could be
observed also in the form of a quasi-periodic X-ray
luminosity~\cite{van_der_klis00}.

	Thirdly, and maybe most importantly, because of their toroidal
topology, the relativistic tori considered here have intrinsically large
mass quadrupoles and if the latter are induced to change rapidly as a
consequence of perturbations or instabilities, large amounts of
gravitational waves could be emitted and possibly
detected~\cite{vp01,zrf03}.

	The plan of this paper is as follows: in Sections 2 and 3~I will
first discuss briefly how stationary, non-Keplerian fluid configurations
can be constructed and the numerical setup necessary to study their
evolution in time. Section 4 is dedicated to a discussion of how to
account for the changes in the spacetime geometry without an explicit
solution of the Einstein equations, while in Section 5 I present the
family of perturbations imposed on the equilibrium models. The numerical
results are finally presented in Section 6, while the conclusions and the
outlook for future developments are summarised in Section 7. Throughout,
I will use a space-like signature $(-,+,+,+)$ and a system of geometrized
units in which $G = c = 1$. The unit of length is chosen to be the
gravitational radius of the black hole, $r_{\rm g} \equiv G M/c^2$, where
$M$ is the mass of the black hole. When useful, however, cgs units will
be reported for clarity. Greek indices are taken to run from 0 to 3.

%=======================================================
\section{Stationary configurations}
\label{analytic}
%=======================================================

	Stationary toroidal fluid configurations in a curved spacetime
can be calculated after solving the Euler equations for a fluid in
circular and non-Keplerian orbits. To this scope I consider a perfect
fluid described by the stress-energy tensor
\begin{equation}
\label{stress-tensor}
T^{\mu\nu}\equiv 
	 \rho h u^\mu u^\nu+p g^{\mu\nu} \ ,
\end{equation}
where $g^{\mu\nu}$ are the coefficients of the metric and are chosen to
be those of a Schwarzschild spacetime in spherical coordinates
$(t,r,\theta,\phi)$.  Here $p$, $\rho$, and $h$ are the isotropic
pressure, the rest-mass density, and the specific enthalpy,
respectively. In the following the fluid will be modeled as ideal with a
polytropic equation of state $p=\kappa \rho^\gamma$, where $\kappa$ is
the polytropic constant and $\gamma$ the adiabatic index.  The motion of
the fluid is non-geodesic with four-velocity $u^{\alpha} =
u^t(1,0,0,\Omega)$, where $\Omega = \Omega(r,\theta) \equiv u^{\phi}/u^t$
is the coordinate angular velocity as observed from infinity.

	Under these assumptions, the equations of motion for the fluid
can be generically written as $h^i_{\ \nu} \nabla_{\mu} T^{\mu \nu}=0$
where $h_{\mu \nu} \equiv g_{\mu \nu} + u_{\mu}u_{\nu}$ is the projector
tensor orthogonal to ${\bf u}$ and $\nabla$ the covariant derivative in
the Schwarzschild spacetime. Enforcing the conditions of hydrostatic
equilibrium and of axisymmetry (i.e. $\partial_t=0=\partial_{\phi}$)
simplifies the hydrodynamical equations considerably. Furthermore, if the
contributions coming from the self-gravity of the torus can be neglected,
the relativistic hydrodynamics equations reduce to Bernoulli-type
equations
\begin{equation}
\label{bernoulli}
\frac{\nabla_i p}{e+p} = - \nabla_i W +
        \frac{\Omega \nabla_i \ell}{1- \Omega \ell} \ ,
\end{equation}
where $i=r,\theta$ and $W = W(r,\theta) \equiv \ln(u_t)$. The local
extrema of the potential $W = W(r,\theta) \equiv \ln(u_t)$ on the
equatorial plane identify the positions of the two Keplerian points of
the solution. At these locations an orbiting fluid element would not
experience any acceleration coming from pressure gradients, the
centrifugal force balancing the gravitational force exactly.
\begin{figure}[htb]
\centering \includegraphics[width=10.5cm]{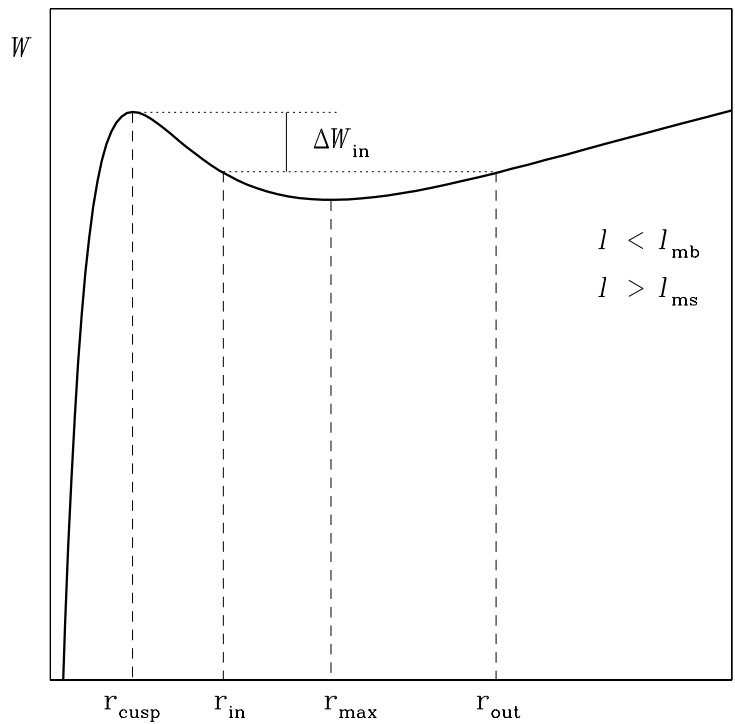} 
\caption{ Potential $W=\ln(u_t)$ on the equatorial plane for a constant
specific angular momentum distribution with $\ell_{\rm ms} < \ell <
\ell_{\rm mb}$. The two extrema mark the position of the Keplerian
points, namely the cusp, $r_{{\rm cusp}}$, and the maximum rest-mass
density, $r_{{\rm max}}$, of the disc. The inner edge $r_{{\rm in}}$ can
be chosen anywhere between $r_{{\rm cusp}}$ and $r_{{\rm max}}$. The
energy gap $\Delta W_{{\rm in}}=W_{{\rm in}}-W_{{\rm cusp}}$ is used to
fix $r_{{\rm in}}$. }
\label{punti}
\end{figure}
These points correspond to the positions of the cusp, $r_{\rm{cusp}}$,
and of the maximum rest-mass density (i.e. the ``centre'' of the torus)
$r_{\rm max}$ (see Figure~\ref{punti}).  While a detailed classification
of the possible models can be found in the literature \cite{abramowicz78,
font02}, here I simply recall that different models for the tori will be
determined after different radial distributions for the specific angular
momentum $\ell \equiv -u_{\phi} /u_t =\ell(r \sin\theta)$ (i.e. the
angular momentum per unit energy) have been chosen.

	An obvious and widely discussed choice for the specific angular
momentum is the one in which the specific angular momentum is assumed to
be constant inside the torus and in the range between the specific
angular momentum at the marginally stable orbit $\ell_{\rm ms} =
3\sqrt{6}/3 \simeq 3.67$ and the one at the marginally bound orbit
$\ell_{\rm mb} = 4$, to yield a disc of finite size. In this case then,
the second term on the right-hand-side of equation (\ref{bernoulli}) is
zero and the Euler equations for a polytrope can be integrated
analytically to yield the rest-mass density distribution inside the torus
as
\begin{eqnarray}
\label{density}
\rho(r,\theta) = \left(\frac{\gamma-1}{\kappa
	\gamma}\right)^{1/(\gamma-1)} 	
	\left[\exp({W_{\rm in}-W})-1\right]^{1/(\gamma-1)} \ ,
\end{eqnarray}
where $W_{\rm in} \equiv W(r_{\rm in},\pi/2)$, $r_{\rm in}$ marking the
inner edge of the torus. The models described so far represent the
fiducial background fluid configurations.

%=======================================================
\section{Numerical setup}
\label{numerical}
%=======================================================

	As shown by \cite{marti91}, the general relativistic Euler
equations can be written as a hyperbolic system of conservation laws,
which take the form \cite{banyuls97}
\begin{eqnarray}
\label{fcf}
\frac{\partial {\bf U}}{\partial t} +
	\frac{\partial [\alpha {\bf F}^{r}]}{\partial r} +
	\frac{\partial [\alpha {\bf F}^{\theta}]}{\partial \theta}
	 = {\bf S} \ ,
\label{system}
\end{eqnarray}
where $\alpha \equiv \sqrt{-g_{00}}$ is the lapse function of the
Schwarzschild metric. Here, ${\bf U}=(D, S_r, S_{\theta}, S_{\phi})$ is
the state-vector of the ``conserved'' variables, namely the relativistic
rest-mass density and the momenta in the coordinate directions,
respectively. All of the quantities are as measured by an Eulerian
observer in the Schwarzschild metric. The other vectors ${\bf F}^{i}$ and
${\bf S}$ appearing in (\ref{fcf}) represent the fluxes and sources of
the evolved quantities, respectively, with the latter being entirely due
to the curvature of spacetime.  Explicit expressions of all these terms
can be found in \cite{font98a}.

	The conservative nature of the system (\ref{system}) is the key
ingredient for the use of High-Resolution Shock-Capturing (HRSC)
numerical methods which assure high-order accuracy, sharp resolution of
discontinuities and absence of spurious oscillations everywhere in the
solution, even at relativistic regimes. Such methods, which are also
referred to as ``Godunov-type'' methods, have been employed in general
relativistic calculations only rather recently but have shown their
ability to yield results of unprecedented
accuracy~\cite{font-goodale02}. Within these methods, the time evolution
of the discretized data from a timelevel $n$ to the subsequent one $n+1$
is made according to the following numerical scheme
\vbox{
\begin{eqnarray}
\label{godunov}	
{\bf U}_{i,j}^{n+1} = {\bf U}_{i,j}^{n}
	-  \frac{\Delta t}{\Delta r}
	\left(\widehat{{\bf F}}^r_{i+1/2,j}-
	      \widehat{{\bf F}}^r_{i-1/2,j}\right)
	- \frac{\Delta t}{\Delta \theta}
    	\left(\widehat{{\bf F}}^{\theta}_{i,j+1/2}-
  	      \widehat{{\bf F}}^{\theta}_{i,j-1/2}\right)
	+ \Delta t \,\, {\bf S}_{i,j} \ ,
\end{eqnarray}}
\noindent
where the subscripts $i,j$ refer to spatial ($r,\theta$) gridpoints, so
that ${\bf U}_{i,j}^n \equiv {\bf U}(r_i, \theta_j, t^n)$.  The
inter-cell numerical fluxes, $\widehat{{\bf F}}^{r}_{i \pm 1/2,j}$ and
$\widehat{{\bf F}}^{\theta}_{i,j \pm 1/2}$, are computed using Marquina's
approximate Riemann solver~\cite{dm96}. A piecewise-linear cell
reconstruction procedure provides second-order accuracy in space, while
the same order in time is obtained with a two-step second-order
Runge-Kutta scheme applied to evolution scheme in expression
(\ref{godunov}).

	The computational grid, consisting typically of $N_r \times
N_{\theta}=250 \times 84$ zones in the radial and angular directions,
respectively, covers a computational domain extending from $r_{\rm
{min}}=2.1$ to $r_{\rm {max}}=30$ and from $0$ to $\pi$. Moreover, in
order to achieve the best resolution in the vicinity of the cusp, the
radial grid is logarithmically spaced in terms of a tortoise coordinate
$r_* = r + 2M \ln (r/2M -1 )$, with the maximum radial resolution at the
innermost grid being $\Delta r=6\times 10^{-4}$.

%%%%%%%%%%%%%%%%%%%%%%%%%%%%%%%%%%%%%%%%%%%%%%%%%%%%%%%%%%%%%%%%%%%%%%%
\section{Evolution of the spacetime}
\label{ste}
%%%%%%%%%%%%%%%%%%%%%%%%%%%%%%%%%%%%%%%%%%%%%%%%%%%%%%%%%%%%%%%%%%%%%%%

	In order to assess the occurrence of the runaway instability, the
changes in the spacetime must be taken into account. Ideally, this would
require the solution of the Einstein equations as well as those of
relativistic hydrodynamics. Besides the intrinsic difficulty of this
problem and despite the fact that a number of numerical codes are being
developed for this purpose, the need of high resolution is still the
major obstacle to the goal. 

	Here, to avoid the solution of the full Einstein equations and
yet simulate the evolution of the spacetime, a more pragmatic approach
(already suggested by~\cite{font02}) has been chosen. In practice, the
metric terms are updated from a given timestep to the following one
according to the amount of matter which has accreted into the hole, i.e.
\begin{equation}
\label{met_up}
g_{\mu \nu}(r,M^{n}) \longrightarrow {\tilde
        g}_{\mu \nu}(r,M^{n+1}) \ ,
\end{equation}
where ${\tilde g}_{\mu \nu}$ describes the spacetime at the timelevel
$n+1$ and $M^{n+1}$ is the ``new'' mass of the black hole as computed in
terms of the accretion rate at the inner zone of the grid $\dot
m^n(r_{\rm min})$
\begin{equation}
M^{n+1}=M^n + \Delta t \ \dot m^n(r_{\rm min}) \ ,
\end{equation}
where
\begin{equation}
\dot{m}(r_{\rm min})\equiv - 2\pi\int_0^{\pi} \sqrt{-g} D v^r d\theta
        \Big\vert_{r_{\rm min}}\ . 
\end{equation}

	Simulations employing the time evolution of the metric
(\ref{met_up}) will be referred to as simulations with a {\it dynamical
spacetime} to distinguish them from those simulations in which the black
hole mass is not evolved, $M^{n+1}=M^n$, and that will be referred to as
with a {\it fixed spacetime}.

	Of course, the approach sketched in equation (\ref{met_up}) is
only an approximation which does not account for the transfer of angular
momentum from the torus to the black hole and that prevents from the
calculation of the response of the black hole to the accreted mass and
the corresponding emission of gravitational radiation.  Nevertheless, it
can be considered a satisfactory one when the tori are not very massive
and the rest-mass accretion rates are small.

%=======================================================
\section{Initial data}
\label{initialdata}
%=======================================================

	In previous simulations of an accreting torus over a fixed
\cite{igumenshchev97} or over an evolving \cite{font02} spacetime, a
deviation from the stationary solution was obtained by artificially
expanding the torus by a small amount to overcome the potential barrier,
thus setting the initial configuration out of equilibrium. The quantity
controlling the deviation from equilibrium is the potential jump on the
equatorial plane between the inner edge of the torus and the position of
the cusp, and it is defined as $\Delta W_{\rm in} \equiv W_{\rm {in}} -
W_{\rm{cusp}}$ (see Figure~\ref{punti}).  Clearly, when $\Delta W_{\rm
in}>0$, the torus overflows its Roche lobe and accretion onto the black
hole is not only possible, but rather unavoidable.

	In calculations presented here, instead, all of the models have
potential barriers $\Delta W_{\rm in}\leq 0$, and the departure from the
equilibrium solution is obtained by introducing a perturbation in the
radial velocity field, which is chosen as a homologous parametrization of
the spherical accretion solution in a Schwarzschild spacetime, i.e. the
Michel solution~\cite{michel72}. Using $\eta$ to parametrize the
strength of the perturbation, the initial radial (covariant) component of
the three-velocity has been specified as
\begin{equation}
\label{vpert}
v_r = \eta (v_{r})_{_{\rm Michel}} \ ,
\end{equation}
where $(v_{r})_{_{\rm Michel}}$ is the radial component of the velocity
in the Michel solution. 

	Doing so provides the possibility of testing how sensitive the
runaway instability is on the specific initial conditions chosen and, in
particular, on the assumption that the torus is already overflowing its
Roche lobe. Furthermore, by having a nonzero initial radial velocity in
addition to the orbital one, these initial conditions are more consistent
with the results of a series of Newtonian simulations in which a torus is
produced as a result of the dynamical merging of two neutron
stars~\cite{ruffert99}. In these simulations, in fact, the torus is seen
to maintain an averaged inflow velocity in the central region of the
order of $\sim 3\times 10^{-3}c$.  The values of the parameter $\eta$
have therefore been chosen such as to be compatible with the estimates
provided by \cite{ruffert99} and are in the range $[0.001, 0.06]$.

	A small but representative sample of the equilibrium models
considered in these simulations is reported in Table~\ref{tab1}.

\begin{table*}
\begin{center}
\footnotesize\rm
\caption{Main properties of the constant angular momentum relativistic
tori used in the numerical calculations. From left to right the columns
report: the name of the model, the torus-to-hole mass ratio $M_{\rm
t}/M$, the polytropic constant $\kappa$, the density at the ``centre'' of
the torus and the average density of each model, respectively, both in
cgs units. All of the models share the same mass for the black hole,
$M=2.5M_{\odot}$, and the adiabatic index, $\gamma=4/3$. The inner and
outer radii, the radial position of the cusp and of the ``centre'' are
also the same for these models, and are given by $r_{\rm in}=4.576$,
$r_{\rm out}=15.89$, $r_{\rm cusp}=4.57$, $r_{\rm max}=8.35$,
respectively, in units of $r_{\rm g} \equiv G M/c^2$.\bigskip}
\label{tab1}
\begin{tabular}{ccccc}
\topline
~ & & & & \\ 
Model   & $M_{\rm t}/M$    & $\kappa$ (cgs) & $\rho_{\rm max}$ (cgs)
	& $\langle\rho\rangle$  (cgs)\\
~ & & & & \\ 
\midline
~ & & & & \\ 
(a)   & 1.00   & 4.46${\times} 10^{13}$ &1.14${\times} 10^{14}$ &4.72${\times}
	10^{12}$  \\
(b)   & 0.50   & 5.62${\times} 10^{13}$ &5.72${\times} 10^{13}$ &2.36${\times}
	10^{12}$   \\
(c)   & 0.10   & 0.96${\times} 10^{14}$ &1.14${\times} 10^{13}$ &4.73${\times}
	10^{11}$ \\
(d)   & 0.05   & 1.21${\times} 10^{14}$ &5.73${\times} 10^{12}$ &2.36${\times}
	10^{11}$ \\
~ & & & & \\ 
\topline
\end{tabular}
\end{center}
\end{table*}

%=======================================================
\section{Numerical results}
\label{results}
%=======================================================

%=======================================================
\subsection{Runaway instability}
\label{runaway}
%=======================================================

	Simulations performed in a dynamical spacetime (in the sense
explained in Section~\ref{ste}) and using the initial conditions
discussed in Section~\ref{initialdata} have confirmed what found by
\cite{font02}, namely the occurrence of the runaway instability for
constant angular momentum configurations. Moreover, the simulations have
highlighted that Roche lobe overflowing is not a necessary condition for
the development of the instability, which can therefore take place under
a variety of different initial conditions as long as the distribution of
specific angular momentum is maintained constant.

\begin{figure}[htb]
\centering \includegraphics[width=10.5cm]{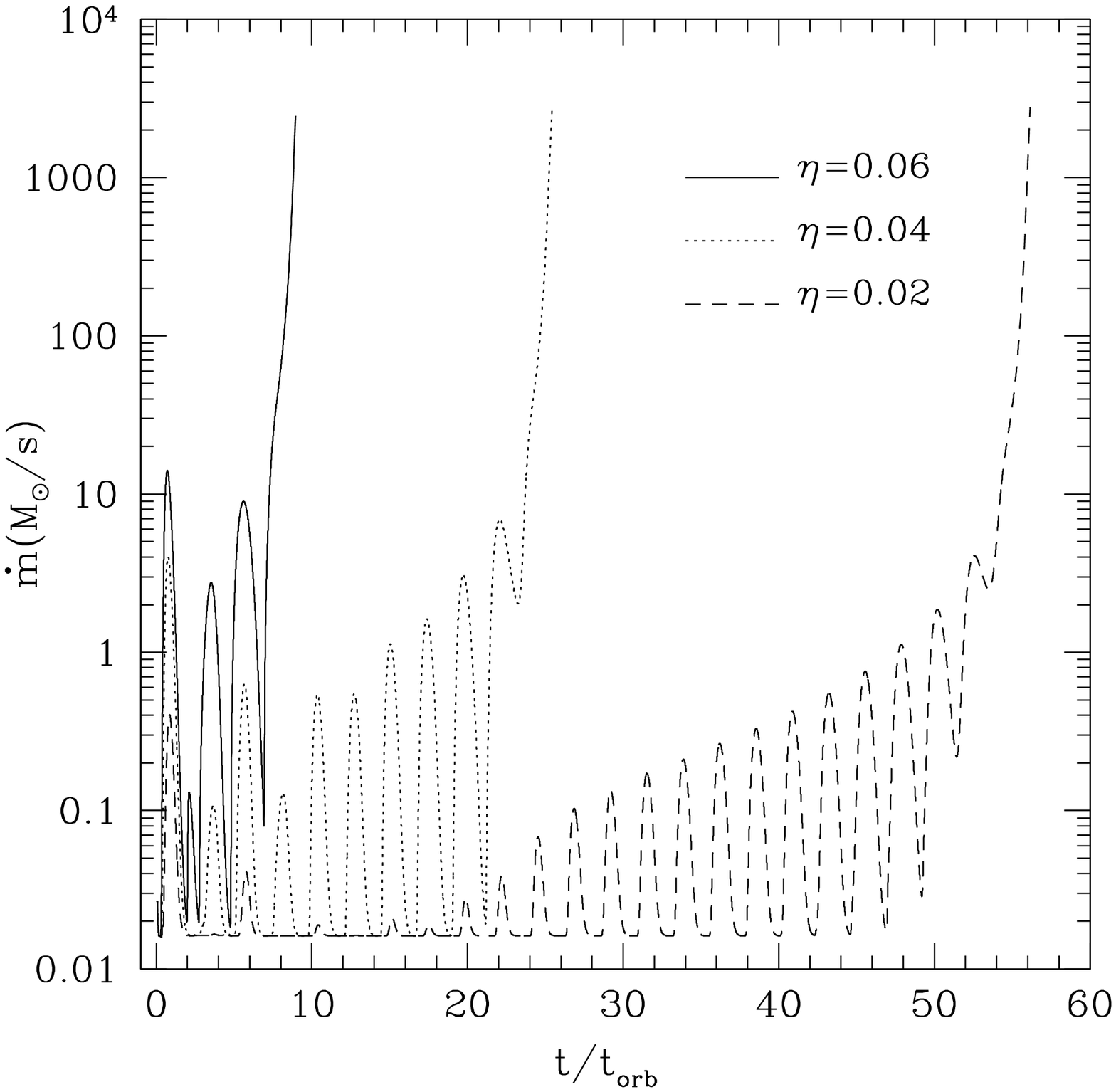} 
\caption{Evolution of the rest-mass accretion rate for model (a) and for
different values of $\eta$ when the spacetime is allowed to vary (i.e. a
dynamical spacetime). The time is expressed in terms of the timescale for
the ``centre'' of the toroidal neutron star to perform an orbit. Note that
the onset and development of the instability is accompanied by periodic
oscillations.}
\label{fig3}
\end{figure}

	Figure~\ref{fig3} shows the evolution of the rest-mass accretion
rate for model $\rm (a)$ in a dynamical spacetime (see Table~\ref{tab1})
and for three different values of initial velocity perturbation,
$\eta$. The time is expressed in terms of the orbital period $t_{\rm orb}
\equiv 2\pi/\Omega_{\rm max}$ for a particle orbiting at the radial
position of the maximum density in the torus, and which is $t_{\rm
orb}=1.87\ {\rm ms}$ for all of the models in Table~\ref{tab1}.  

	As it is apparent from the figure, the accretion rate grows
exponentially after an amount of time which is shorter for larger values
of the initial velocity perturbation. Furthermore, the secular growth in
the rest-mass accretion rate is accompanied by regular, periodic
oscillations. As a consequence of the introduction of the initial
perturbation, in fact, the torus experiences a series of compressions and
expansions during which a considerable amount of matter falls towards the
black hole. In the half-period of oscillation leading to the compression
wave, in particular, the inner edge of the torus moves towards the black
hole and a certain amount of matter can overcome the potential barrier,
thus producing the local maxima in the accretion rate which are visible
in Figure~\ref{fig3}. Note also that the oscillations are not at constant
amplitude, but have increasingly large amplitudes produced by the
development of the instability. This superposition of the onset of the
instability and of the response to the perturbation prevents a direct
analysis of the periodic component; this difficulty, however, can be
overcome simply as will be discussed in Section~\ref{quasi}.

	The calculations during the final stages of the instability
become increasingly difficult because of the dramatical changes in the
hydrodynamical quantities (note the exponential increases in
Figure~\ref{fig3}). In such conditions, Courant factors as small as
$0.01$ must be taken in order to avoid the crashing of the code. Soon
after the accretion rate has reached its maximum, it drops rapidly to
very small values as a consequence of the fact that $98\%$ of the initial
mass of the torus has accreted onto the hole by then.

	When models with smaller mass ratios $M_{\rm t}/M$ are evolved
[i.e. models (b), (c) and (d) in Table~\ref{tab1}] the accretion rate is
generally lower and the amount of matter accreted can be very small even
over several tens of dynamical timescales. In these conditions the
timescales for the onset and development of the instability are
progressively larger, but all of the qualitative features discussed so
far apply unmodified (see \cite{zrf03} for further details).

%%%%%%%%%%%%%%%%%%%%%%%%%%%%%%%%%%%%%%%%%%%%%%%%%%%%%%%%%%%%%%%%%%%%%%%
\subsection{Quasi-Periodic Behaviour}
\label{quasi}
%%%%%%%%%%%%%%%%%%%%%%%%%%%%%%%%%%%%%%%%%%%%%%%%%%%%%%%%%%%%%%%%%%%%%%%

	When concentrating on the quasi-periodic response of the torus to
perturbations, the occurrence of the runaway instability represents a
complication to avoid. Since the occurrence of the instability itself is
still under debate and could be suppressed when more generic initial
conditions are considered (e.g. with a non-constant specific angular
momentum distribution \cite{font202}), in the rest of the simulations
presented here the spacetime has been maintained fixed, hence suppressing
the onset of the runaway instability.

\begin{figure}[htb]
\centering
\includegraphics[width=10.5cm]{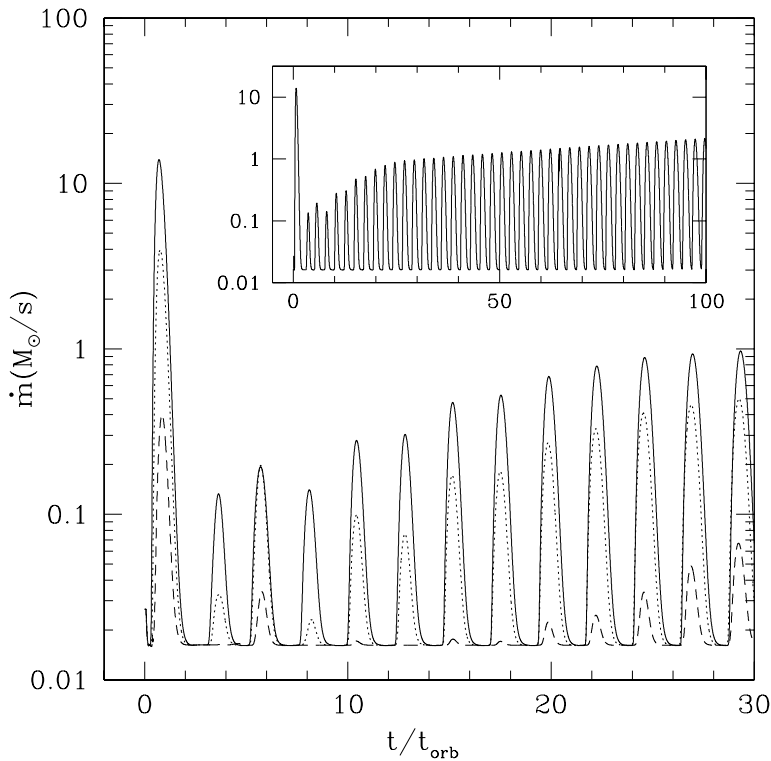}

 \caption{ Rest-mass accretion rate in a fixed-spacetime evolution. The
solid, dotted and dashed lines correspond to $\eta=0.06, 0.04$, and
$0.02$, respectively. The inset shows the case $\eta=0.06$ over a very
long timescale (100 orbital periods).}
\label{fig4}
\end{figure}

	Figure~\ref{fig4} is the same as Figure~\ref{fig3}, but shows the
evolution of the accretion rate during the first 30 orbital periods for a
simulation in which the spacetime is held fixed. As in the case of a
dynamical spacetime, the amplitude of the rest-mass accretion rate shows
a dependence on the strength of the initial perturbation, with ${\dot m}$
becoming progressively larger with increasing $\eta$. In these
simulations, however, the accretion rate never diverges and the pulsating
behaviour can therefore be studied more in detail. Each period of
oscillation in Figure~\ref{fig4} is the result of several tens of
thousands timesteps and the ability to preserve the periodic nature of
the oscillations confirms the ability of HRSC methods in reproducing
subtle hydrodynamical effects with very high precision.  Note also that,
as shown in the inset of Figure~\ref{fig4}, the periodic behaviour is not
altered even when observed over 100 orbital timescales, until the
numerical simulation is stopped or the relativistic torus has been
entirely accreted by the black hole.

\begin{figure}[htb]
\centering
\includegraphics[width=10.5cm]{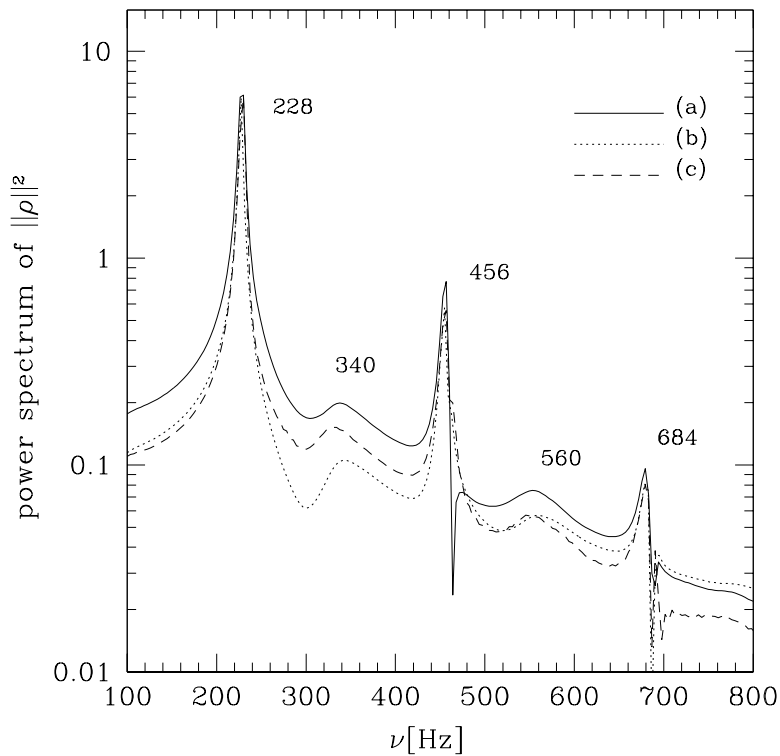}
 \caption{
Power spectrum of the rest-mass accretion rate for models (a), (b), and
(c) of Table~1. The values on the vertical axis have been suitably
normalized to match the power in the fundamental mode and arbitrary units
have been used. }
\label{fig13}
\end{figure}
	
	It is important to underline that the rest-mass accretion rate is
not the only quantity showing a periodic behaviour and, indeed, all of
the fluid variables can be shown to oscillate periodically. However,
before presenting the physical interpretation of the oscillations
detected, it is useful to quantify the quasi-periodicity observed in the
simulations in terms of a Fourier analysis of some relevant
hydrodynamical quantities. As a good representative case,
Figure~\ref{fig13} shows the power spectrum of the $L_2$ norm of the
rest-mass density (i.e. an integral quantity over the numerical grid) for
models (a), (b) and (c) of Table~\ref{tab1}. The Fourier transform has
been calculated with data obtained with an $\eta=0.06$ perturbation and
computed over a time interval going up to $t/t_{\rm orb} \simeq 100$.

	As it is evident from Figure~\ref{fig13}, all of the three power
spectra consist of a fundamental frequency $f_0$ (228 Hz for the models
considered in Figure~\ref{fig13}) and a series of overtones (at 340, 456
Hz, and so on) in a ratio which can be determined to be $2:3:4: \ldots$,
to an accuracy of a few percent. This clearly suggests that the
quasi-periodic response observed is the consequence of fundamental modes
of oscillation of the torus. While reminiscent of epicyclic oscillations,
these modes cannot be interpreted in terms of radial perturbations of
circular Keplerian orbits as the epicyclic frequency is zero for constant
specific angular momentum distributions (This property holds in Newtonian
gravity but has been shown to hold also in a relativistic
regime~\cite{ryz03}.). These oscillations should indeed be interpreted in
terms of {\it ``p modes''} (or inertial-acoustic modes) of the torus.

	To prove this, a detailed linear perturbative analysis has been
performed~\cite{ryz03}, in which it has been shown that $p$ modes in a
general non-Keplerian disc represent the vibrational modes having
pressure gradients as the restoring force, and appear with
eigenfrequencies in the same sequence $2:3:4:\ldots$ revealed here. A
detailed discussion of the results of this perturbative analysis can be
found in~\cite{ryz03} and will not be further discussed here. However, a
much simpler argument can be given to justify the interpretation of the
oscillations detected as $p$ modes. As a by-product of the assumption of
constant specific angular momentum distributions, in fact, the sound
speed is independent of the polytropic constant in these tori. Since the
power spectra reported in Figure~\ref{fig13} refer to models having the
same spatial extensions but differing in the values for the polytropic
constant (and hence in the total rest-mass), they should all possess
identical features if the oscillations are to reflect sound wave
propagation in thick discs. Indeed, this is what shown in
Figure~\ref{fig13}, where all the frequencies differ for less than
$0.1\%$ despite the tori producing them have considerably different
masses and hence different average densities.

	The approach discussed here represents a first step towards a
relativistic disco-seismology analysis for massive and vertically
extended tori in General Relativity. This is in close analogy to what has
been extensively developed for geometrically thin discs~\cite{kato01,
perez97, silbergleit01, rodriguez02} and may serve as a guide to
interpret astrophysical phenomena where thick discs play an important
role.

%%%%%%%%%%%%%%%%%%%%%%%%%%%%%%%%%%%%%%%%%%%%%%%%%%%%%
\subsection{Linear and Nonlinear Regimes}
%%%%%%%%%%%%%%%%%%%%%%%%%%%%%%%%%%%%%%%%%%%%%%%%%%%%%

\begin{figure}[htb]
\centering
\includegraphics[width=10.5cm]{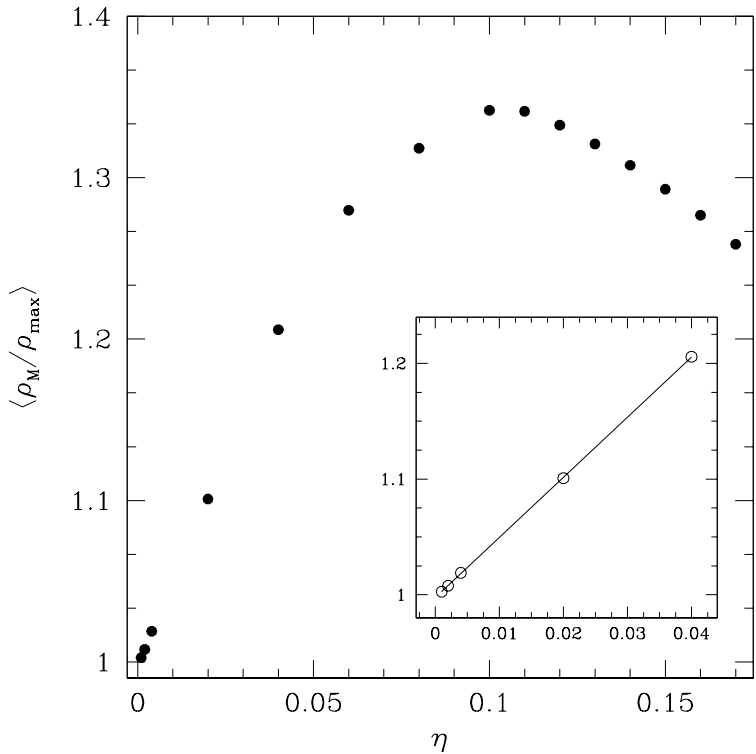}
\caption{Averaged maximum rest-mass density normalized to the central one
in the toroidal neutron star as a function of the perturbation
strength. The small inset shows a magnification of the behaviour in the
linear regime and the solid line shows the good linear fit to the
data. The data refers to model (a) and has been averaged over 10 orbital
timescales.
\label{fig14}}
\end{figure}

	All of the quasi-periodic behaviour discussed so far is the
consequence of the finite size perturbations that have been introduced in
the initial configuration. Within this approach a {\it linear} regime is
expected in which the response of the toroidal neutron star is linearly
proportional to the perturbation introduced, and a {\it nonlinear} regime
when this ceases to be true. The strength of the perturbation which marks
the transition between the two regimes can be estimated from
Figure~\ref{fig14}, which shows the averaged maximum rest-mass density
reached during the oscillations $\rho_{\rm M}$, normalized to the value
at the centre of the toroidal neutron star $\rho_{\rm max}$.

	A rapid look at Figure~\ref{fig14} reveals the presence of both
the linear and nonlinear regimes with the first one being shown magnified
in the inset, and where the solid line shows the good linear fit to the
data. The transition between the two regimes seems to occur for $\eta
\simeq 0.05$, with the nonlinear regime producing maximum amplitudes that
are $\sim 35\%$ larger than the initial one. A careful analysis of the
behaviour of the fluid variables shows that for perturbations with
strength $\eta \gtrsim 0.05$, some of the kinetic energy which is
confined to the lower order modes in the linear regime, tends to be
transferred also to higher order modes. The nonlinear coupling among
different modes and the excitation of higher order overtones is often
encountered in Nature where it serves to redistribute the excess kinetic
energy before the production of shocks. In practice, the nonlinear
coupling deprives of energy the fundamental mode (which is the one
basically represented in Figure~\ref{fig14}) and is therefore responsible
for the decay of $\langle \rho_{\rm M}/\rho_{\rm max}\rangle$ for $\eta
\gtrsim 0.1$. Interestingly, when analysed in terms of the power spectra,
this effect shows a very distinctive behaviour. As the nonlinear
mode-mode coupling becomes effective, the amount of power in the
fundamental mode becomes increasingly smaller as the strength of the
perturbation is increased. At the same time, the conservation of energy
transfers power to the overtones, with the first ones reaching amplitudes
comparable to the fundamental one and with the high order ones becoming
more and more distinct from the background.

	Determining the transition to the nonlinear regime is important
to set an approximate upper limit on the amplitude of the oscillations
and is relevant when estimating the emission of gravitational waves. It
should also be noted that in the parameter range for $\eta$ in which the
calculations have been performed (i.e. $\eta \in [0.001,0.12]$), the peak
frequencies in the power spectra have not shown to depend on the values
used for $\eta$. This is of course consistent with them being fundamental
frequencies (and overtones).

%%%%%%%%%%%%%%%%%%%%%%%%%%%%%%%%%%%%%%%%%%%%%%%%%%%%%
\subsection{Gravitational Wave Emission}
\label{gwe}
%%%%%%%%%%%%%%%%%%%%%%%%%%%%%%%%%%%%%%%%%%%%%%%%%%%%%

	Although the present analysis is not able to calculate in a
self-consistent manner the emission of gravitational waves produced by
the linear and nonlinear dynamics of the torus, it is nevertheless
interesting to attempt a first estimate of this. The rationale is that
during the oscillations the tori undergo large and rapid variations of
their mass quadrupole moments, which are intrinsically large for fluid
configurations with high density and a toroidal topology. As a result,
perturbed toroidal neutron stars undergoing periodic oscillations have
the potential of being strong sources of gravitational waves.

	In the Newtonian quadrupole approximation, the gravitational
waveform $h^{TT}(t)$ observed at a distance $R$ from the source can be
calculated in terms of the quadrupole wave amplitude $A_{20}^{\rm E2} $
(see also~\cite{zwerger97})
\begin{equation}
\label{htt}
h^{TT}(t) = F_+ \left(\frac{1}{8}
	  \sqrt{\frac{15}{\pi}}\right) \frac{A_{20}^{\rm E2}(t-R)}{R} \ ,
\end{equation}
where $F_+=F_+(R,\theta,\phi)$, the detector's beam pattern function, is
assumed to be optimal, i.e. $F_+=1$.  The wave amplitude $A_{20}^{\rm
E2}$ in Eq. (\ref{htt}) is the second time derivative of the mass
quadrupole moment $A_{20}^{\rm E2} = d^2 I/dt^2$, where $I$ is defined as
\begin{equation}
\label{mq}
I\equiv \int \rho \left(\frac{3}{2}z^2 - \frac{1}{2}\right)r^4 dr dz \ ,
\end{equation}
with $z \equiv \cos\theta$.  

\begin{figure}[htb]
\centering \includegraphics[width=10.5cm]{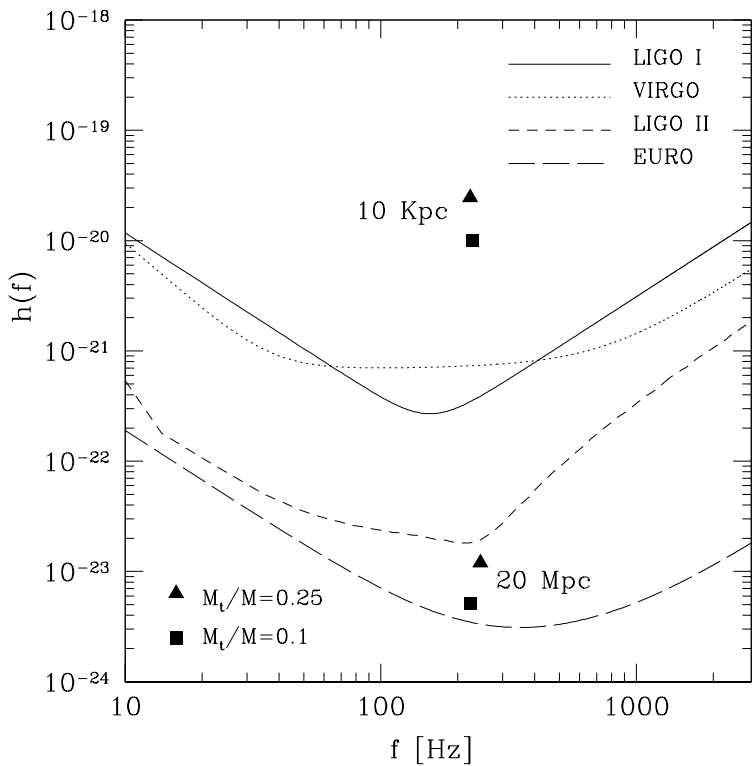} \caption{
Characteristic wave amplitudes for a perturbed toroidal neutron star with
$\eta=0.04$. These amplitudes have been computed using the strain noise
estimated for LIGO I for a source located at a distance of 10 Kpc and for
LIGO II for a source at a distance of 20 Mpc. The numbers can also be
compared with the sensitivity curves of VIRGO and EURO that are similar
in this frequency range. Different points refer to different mass ratios,
with triangles indicating $M_{\rm t}/M=0.25$ and squares $M_{\rm
t}/M=0.1$.}
\label{fig11}
\end{figure}

	Within this Newtonian framework and using the results of the
numerical calculations to estimate the gravitational wave amplitude, it
is then possible to derive a phenomenological expression for the
gravitational waveform that could be expected as a result of the
oscillations induced in the torus. In particular, a source with mass
$M_{\rm t}$ and located somewhere in the Galaxy (i.e. at a distance $R=10
$~Kpc), would produce a transverse traceless gravitational wave amplitude
given by
\begin{equation}
\label{h_phenom}
h^{TT} \simeq 2.2 \times 10^{-21} \left(\frac{\eta}{0.04}\right)
	\left(\frac{M_{\rm t}}{0.1\ M_{2.5}}\right)
	\left(\frac{10\; {\rm Kpc}}{R}\right) \ , 
\end{equation}
where $M_{2.5}\equiv M/(2.5\ M_{\odot})$. Expression~(\ref{h_phenom})
shows that, already in the linear regime, a non-negligible gravitational
wave amplitude can be produced by an oscillating toroidal neutron star
orbiting around a black hole. This amplitude is comparable with the
average gravitational wave amplitude computed in the case of core
collapse in a supernova explosion \cite{zwerger97,dimmel02} and can
become stronger for larger perturbations or masses in the torus.

	In order to quantify the detectability of relativistic tori by
present and planned the interferometric detectors, it is necessary to
compute the {\it characteristic} gravitational wave frequency and
amplitude, as well as the corresponding signal-to-noise ratio for these
detectors. More specifically, these quantities have been evaluated
following the procedure discussed in~\cite{zrf03} and use, for a
``realistic'' toroidal neutron star model with mass $M_{\rm t}/M = 0.1$
and $\eta=0.04$, a representative timescale $\tau_{\rm life} = 0.18$
s. Longer (shorter) timescales are instead assumed for models with
smaller (larger) initial perturbations (cf. Table~\ref{tab2}).

\begin{table*}
\begin{center}
\footnotesize\rm
\caption{Characteristic properties for the detection of the gravitational
wave signal emitted by a toroidal neutron star. The first two columns
show the mass of the toroidal neutron star normalized to the black hole
one, as well as the strength of the initial perturbation $\eta$. The
following three columns report the signal-to-noise ratio computed for
LIGO I, LIGO II and EURO, assuming a galactic distance (third column) or
an extragalactic one (fourth and fifth columns) . The last column reports
the characteristic lifetime for the existence of relativistic tori that
are unstable to the runaway instability. Of course, a toroidal neutron
star may survive longer than this timescale should the runaway
instability not occur.\bigskip}
\label{tab2}
\begin{tabular}{ccccccc}
\topline
$M_{\rm t}/M$
       & $\eta$ & $S/N$ LIGO I & $S/N$ LIGO II & $S/N$ EURO
       & $\tau_{\rm life}$ (s) \\
       &          & (10 Kpc) & (20 Mpc)  & (20 Mpc) &   \\
\midline
$0.10$ & $0.06$ & $23.2$ & $0.1$  & $1.8$ & 0.14 	\\
$0.10$ & $0.04$ & $19.9$ & $0.1$  & $1.5$ & 0.18 	\\
$0.10$ & $0.02$ & $11.9$ & $<0.1$ & $0.9$ & 0.20 	\\
~							\\
\topline
\end{tabular}
\end{center}
\end{table*}

	Figure~\ref{fig11} and Table~\ref{tab2} show a small but
significant fraction of the parameter space investigated by the numerical
simulations. Figure~\ref{fig11}, in particular, shows the characteristic
wave amplitude for sources located at a distance of $R=10$ Kpc and $R=20$
Mpc, as computed for two different values of the torus-to-hole mass
ratio. These amplitudes have been computed for the expected strain noise
of LIGO I and LIGO II, but the strain curves of VIRGO (that is similar in
this frequency range,~\cite{damour01}) and of EURO~\cite{euro01} have
also been reported for comparison. Interestingly, with small initial
perturbations $(\eta=0.04)$ and mass ratios $(M_{\rm t}/M=0.1)$, the
computed characteristic amplitudes can be above the sensitivity curves of
LIGO I for sources within 10 Kpc and above the sensitivity curve of EURO
for sources within 20 Mpc. Both results suggest that a relativistic torus
oscillating in the Virgo cluster could be detectable by the present and
planned interferometric detectors.

	Table~\ref{tab2} provides a more quantitative measure of the
likelihood of the detection of toroidal neutron stars as sources of
gravitational waves by reporting the signal-to-noise ratios as computed
for LIGO I, LIGO II and EURO, as well as the timescale over which the
signal has been computed.  All of these quantities refer to models with
different initial perturbations and located at either 10 Kpc (for LIGO I)
or 20 Mpc (for LIGO II and EURO). As it is apparent from the table, even
an extragalactic source with a small amount of perturbations could emit a
gravitational wave signal strong enough to be detected by the planned
detector EURO.

%=======================================================
\section{Conclusions}
\label{concl}
%=======================================================

	I have presented general relativistic hydrodynamics simulations
of relativistic axisymmetric tori orbiting around a Schwarzschild black
hole with a constant specific angular momentum. The simulations make use
of HRSC methods and are able to reproduce faithfully both the linear and
the nonlinear response of the relativistic tori. There are three main
aspects of this research that are worth to be highlighted.

	The first one concerns the issue of the runaway instability, a
dynamical instability which is expected to take place when a
geometrically thick disc accretes matter onto a black hole through the
cusp of its outermost equipotential surface. The calculations reported
here for a variety of different models show that, at least for constant
specific angular momentum tori whose self-gravity is neglected, the
runaway instability represents a robust, unavoidable feature of their
dynamics. Furthermore, these simulations clarify that an initial Roche
lobe overflow is not a necessary condition for the development of the
instability.

	The second aspect concerns the dynamical response of these
relativistic tori to the perturbations that will be present when these
objects form.  Upon the introduction of suitably parametrized
perturbations, the tori have shown a regular oscillatory behaviour
resulting both in a quasi-periodic variation of the rest-mass accretion
rate as well as of the rest-mass distribution. This response is to be
interpreted in terms of the excitation of $p$ modes having pressure
gradients as restoring forces and appearing with eigenfrequencies in the
sequence $2:3:4:\ldots$.

	Finally, as a consequence of the excitation of oscillations, the
mass quadrupoles are induced to change rapidly and intense gravitational
radiation is thus produced. Estimates made within the Newtonian
quadrupole approximation have shown that strong gravitational waves can
be produced during the short lifetime of these tori.  In particular, the
gravitational radiation emitted by these sources is comparable or larger
than the one that is expected during the gravitational collapse of a
stellar iron core, with a rate of detectable events which could also be
larger given the variety of physical scenarios leading to the formation
of a massive torus orbiting a black hole.

\section*{Acknowledgments}

The work reported here is based on the results presented in~\cite{zrf03}
and ~\cite{ryz03} and has been made in collaboration with Toni Font,
Shin Yoshida and Olindo Zanotti, who are gratefully acknowledged.
Financial support for this research has been provided by the MIUR and by
the EU Network Programme (Research Training Network Contract
HPRN-CT-2000-00137). The computations were performed on the Beowulf
Cluster for numerical relativity {\it ``Albert100''}, at the University
of Parma.

\end{document}